\documentclass[times,dvipdfmx]{akari2017} 

\newcommand     \mum    {\,\mu{\rm m}}  

\usepackage{natbib}
\usepackage{color}
\usepackage{hyperref} 
\hypersetup{colorlinks=true,citecolor=blue}

\def \bea {\begin{eqnarray}}
\def \ena {\end{eqnarray}}

\def 	\bE	{{\bf E}}

\def	\bJ	{{\bf J}}

\def    \bmu    {{\hbox{\boldsym\char'026}}}	

\def    \bomega {{\hbox{\boldsym\char'041}}}	

\def	\H	{{\rm H}}

\def	\rot	{{\rm rot}}

\def	\Bar	{{\rm Bar}}

\def	\zhat		{\hat{\bf z}}
\def	\ahat		{\hat{\bf a}}

\def    \Bv     	{\bf  B}

\font\mib=cmmib10

\def\bomega{\hbox{\mib\char"21}}

\def\bmu{\hbox{\mib\char"16}}

\shorttitle{Polarized PAH emission}
\shortauthors{Thiem Hoang}

\begin{document}

\title{Polarization of Infrared Emission from Polycyclic Aromatic Hydrocarbons} 



\correspondingauthor{Thiem Hoang}
\email{thiemhoang@kasi.re.kr}

\author{Thiem Hoang} 
\affiliation{Korea Astronomy and Space Science Institute, Daejeon 34055, Korea; thiemhoang@kasi.re.kr} 
\affiliation{Korea University of Science and Technology, Daejeon, 34113, Korea}



\begin{abstract}
Polarized infrared emission from polycyclic aromatic hydrocarbons (PAHs) is important for testing the basic physics of alignment of ultrasmall grains and potentially offers a new way to trace magnetic fields. In this paper, a new model of polarized PAH emission is presented, taking into account the effect of PAH alignment with the magnetic field. The polarization level of PAH emission features, for the different phases of the diffuse interstellar medium (ISM) is discussed. We find that negatively charged smallest PAHs in the reflection nebula can be excited to slightly suprathermal rotation due to enhanced ion collisional excitation, which enhances the degree of PAH alignment and the polarization level of PAH emission. The polarization level and polarization angle predicted by our model including PAH alignment are supported by the first detection of the polarization of $p\sim 2\%$ at 11.3 $\mu$m PAH feature from MWC 1080 nebula by \cite{Zhang:2017ea}. The theoretical and observational progress reveals that PAHs can be aligned with the magnetic field, resulting in a moderate polarization level of spinning dust emission. Polarized infrared PAH polarization would be useful for tracing the magnetic fields.
\end{abstract}



\setcounter{page}{1}

\section{Introduction}   
Polycyclic aromatic hydrocarbons (PAHs) is an important dust component of the interstellar medium (ISM, see \citealt{2008ARA&A..46..289T} for a review). PAH molecules are planar structures, consisting of carbon hexagonal rings and hydrogen atoms attached to their edge via valence bonds. PAHs are the leading carrier of the 2175\AA~extinction bump. Upon absorbing ultraviolet (UV) photons, PAHs reemit radiation in mid-infrared features, including 3.3, 6.2, 7.7, 8.6, 11.3, and 17 $\mu$m, due to vibrational transitions (\citealt{1984A&A...137L...5L}; \citealt{1985ApJ...290L..25A}). Recent IR observations by {\it Spitzer} and {\it AKARI} (\citealt{2007PASJ...59S.401O}) provide a wealth of data on PAH emission from galaxies.

Rapidly spinning PAHs also emit rotational radiation in microwaves via a new mechanism, so-called spinning dust (\citealt{1998ApJ...508..157D}; \citealt{Hoang:2010jy}). The latter is the most likely origin of anomalous microwave emission (AME) that contaminates Cosmic Microwave Background (CMB) radiation  (\citealt{Kogut:1996p5293}; \citealt{Leitch:1997p7359}). 

CMB experiments aiming to detect primordial gravitational waves through B-mode polarization face great challenges from polarized Galactic foregrounds (\citealt{Ade:2015ee}; \citealt{2016A&A...586A.133P}), including thermal dust emission and anomalous microwave emission (AME). Modern understanding shows that the AME is most likely produced by rapidly spinning nanoparticles, including PAHs, silicate (\citealt{2016ApJ...824...18H}; \citealt{2017ApJ...836..179H}) and iron nanoparticles \citep{2016ApJ...821...91H}. \footnote{Throughout this paper, dust grains refer to grains above 10 nm (or $100$\AA), while nanoparticles refer to ultrasmall grains smaller than ~10 nm.} Thus, polarization of PAH emission can shed light on alignment of nanoparticles, which will provide constraints on AME polarization. 

The polarization of PAH emission features, however, depends both on the internal alignment of the grain axis of major inertia $\ahat_{1}$ with $\bJ$ and the external alignment of $\bJ$ with the magnetic field $\Bv$. Modeling polarized PAH emission is thus complicated because PAH emission only occurs during a short time interval following UV photon absorption, while the dynamical (e.g., rotational damping and grain alignment) timescales are much longer. 
The first model of polarized PAH emission is presented by \cite{1988prco.book..769L} (hereafter L88) where the author noticed that internal alignment can produce polarized emission when PAHs being illuminated anisotropically by UV photons. L88 estimate the polarization of $\sim 0.9\%$ for the 3.3 $\mu$m and of $\sim-2.1\%$ for 11.3$\mum$ features, respectively. \cite{Sironi:2009p5558} (hereafter SD09) revisited the L88's model by considering the realistic rotational dynamics of PAHs and found much lower polarization fractions, i.e., of $\sim 0.06\%$ for 3.3$\mum$ and $\sim -0.53\%$ for 11.3$\mum$, for the conditions of Orion Bar considered by L88. 

Both L88 and SD09 models assumed randomly oriented grain angular momentum in the space, which results in the underestimates of the polarization level because PAHs are expected to be partially aligned due to paramagnetic resonance relaxation (\citealt{2000ApJ...536L..15L}). \cite{2017ApJ...838..112H} presents a new model of polarized emission by incorporating the effect of the partial alignment of $\bJ$ with $\Bv$ due to resonance paramagnetic relaxation mechanism and presented. The author also used latest progress in grain rotational dynamics (e.g., grain wobbling, anisotropic damping and excitation by IR emission) achieved in our previous works (\citealt{{Hoang:2010jy},{2011ApJ...741...87H}}).

The structure of our paper is as follows. We first discuss relevant physics of PAHs and alignment mechanisms in Section \ref{sec:PAHalign}. In Section \ref{sec:model}, we describe the coordinate systems, numerical methods for calculations of polarization degree by including partial alignment of the angular momentum and the magnetic field and numerical results. We discuss the implications of our obtained results in Section \ref{sec:discus}. 

\section{Physics of PAH alignment}\label{sec:PAHalign}
In this section, we will briefly describe the basic physics of alignment of spinning PAHs and nanoparticles. The alignment of PAHs requires the existence of a magnetic moment in the spinning nanoparticle.
\subsection{Magnetic properties of PAHs}
Ideal PAHs are expected to have rather low paramagnetic susceptibility due to H nuclear spin \citep{Jones:1967p2924}. However, astrophysical PAHs are likely magnetized thanks to the presence of free radicals, paramagnetic carbon rings, or adsorption of ions (see \citealt{2000ApJ...536L..15L}). Modern understanding in graphene magnetism found that graphene can be magnetized due to C vacancy or H adorbsion on the graphene surface (\citealt{2007PhRvB..75l5408Y}; \citealt{2004PhRvL..93r7202L}), which is detected in a recent experiment (\citealt{2016Sci...352..437G}). In the ISM, the defects of PAHs can be triggered by bombardment of cosmic rays. All together, astrophysical PAHs are likely paramagnetic.  Paramagnetic grains acquire the instantaneous magnetic moment along the grain angular velocity $\bomega$ due to its rotation (\citealt{Barnett:1915p6353}; \citealt{1976Ap&SS..43..257D}), which induces the coupling of PAHs with the magnetic field through Larmor precession.

\subsection{Internal relaxation and resonance paramagnetic relaxation}

\cite{1979ApJ...231..404P} realized that the precession of $\bomega$ coupled to $\bmu_{\Bar}$ around the grain symmetry axis $\ahat_{1}$ produces a rotating magnetization component within the grain body coordinates. As a result, the grain rotational energy is gradually dissipated into heat until $\bomega$ becomes aligned with $\ahat_{1}$-- an effect that Purcell termed "Barnett relaxation". Internal relaxation (i.e., Barnett, nuclear relaxation, and imperfect elasticity) enables the transfer of grain rotational energy to the vibrational system. Naturally, some vibrational energy can also be transferred to the rotational system \citep{Jones:1967p2924}. For an isolated grain, a small amount of energy gained from the vibrational modes can induce fluctuations of the rotational energy $E_{\rot}$ when the grain angular momentum $\bJ$ is conserved (\citealt{1994MNRAS.268..713L}).

\cite{1951ApJ...114..206D} suggested that a paramagnetic grain rotating with angular velocity $\bomega$ in an external magnetic field $\Bv$ experiences paramagnetic relaxation due to the lag of magnetization, which dissipates the grain rotational energy into heat. This results in the gradual alignment of $\bomega$ and $\bJ$ with the magnetic field until the rotational energy is minimum. For ultrasmall grains, such as PAHs, the classical Davis-Greenstein relaxation is suppressed because the rotation time is shorter than the electron-electron spin relaxation time $\tau_{2}$ (see \citealt{2014ApJ...790....6H}). Yet such nanoparticles can be partially aligned by resonance paramagnetic relaxation that originates from the splitting of the rotational energy \citep{2000ApJ...536L..15L}. Numerical calculations in \cite{2014ApJ...790....6H} showed that PAHs can be aligned by resonance paramagnetic relaxation with the degree $Q_{J}\sim 0.05-0.15$ depending on $T_{0}$ and the magnetic field strength. 

\subsection{Rotational dynamics of PAHs}
The magnetic alignment efficiency of PAHs increases with increasing the rotation rate. The rotation rate of PAHs and nanoparticles depends on various interactions with the gas (neutral, ion) and radiation field. Neutral grain experience direct collisions with neutral gas atoms. Charged grain undergo Coulomb interaction with ion. Neutral grain also interact with passing plasma ion due to the grain electric dipole moment in the ion electric field. Nanoparticles absorb UV photon and reemit IR photons, resulting in the excitation and damping of the rotation (\citealt{1998ApJ...508..157D}; \citealt{Hoang:2010jy}).

\section{New Model of Polarized IR Emision with PAH alignment} \label{sec:model}

\subsection{PAH emission and Polarization}
Let $u,v$ be the two orthogonal directions in the plane of the sky. The coordinate systems used for our modeling are shown in Figure \ref{fig:figures}(left panel).  Let $F_{u,v}^{\|,\perp}$ be the emission flux by a PAH molecule due to in-plane ($\|$) and out-of-plane oscillation ($\perp$) with the electric field $\bE$ in the $\hat{\bf u}$ and $\hat{\bf v}$ directions, respectively (see Figure \ref{fig:figures}(left panel)). Let $I_{u,v}$ be the total emission intensity from the PAH. 

The emission flux $F_{u,v}^{\|,\perp}$ depends on the orientation of the PAH plane with $\bJ$ and the distribution of $\bJ$ with the magnetic field, which are described by the distributions $f_{\rm LTE}(\theta, J)$ and $f_{J}(\bJ)$, respectively. Thus, the total emission intensity is obtained by integrating over these distribution functions:
\bea
I_{\star,w}^{\|,\perp}(\alpha, \psi) &=& \int f_{J}(\bJ)d\bJ\int_{0}^{\pi}f_{0}(\theta_{0},J)d\theta_{0}\int_{0}^{\pi}f_{ir}(\theta, J)d\theta A_{\star}(\beta,\theta_{0})F_{w}^{\|,\perp}(\beta, \varphi,\theta,\alpha),\label{eq:Istar_uv}
\ena
where $w=u,v$, $A_{\star}$ is the cross-section of UV absorption, $f_{0},f_{ir}$, and $F_{w}$ are given in \cite{2017ApJ...838..112H}.

The polarization degree of PAH emission features due to in-plane and out-of-plane oscillations is calculated as the following:
\bea
p^{\|,\perp}(\alpha, \psi)=\frac{I_{u}^{\|,\perp}(\alpha, \psi)-I_{v}^{\|,\perp}(\alpha, \psi)}{I_{u}^{\|,\perp}(\alpha, \psi)+I_{v}^{\|,\perp}(\alpha, \psi)},\label{eq:pol_a}
\ena
where the positive and negative $p$ correspond to the polarization vector along the $\hat{\bf u}$- and $\hat{\bf v}$- direction, respectively (see Figure \ref{fig:figures}(left panel)).

The polarization by a population of PAHs with size distribution $dn/da$ is
\bea
p^{\|,\perp}(\alpha, \psi)=\frac{I_{u}^{\|,\perp}(\alpha, \psi)-I_{v}^{\|,\perp}(\alpha, \psi)}{I_{u}^{\|,\perp}(\alpha, \psi)+I_{v}^{\|,\perp}(\alpha, \psi)},\label{eq:pol}
\ena
where the intensity obtained by integrating over the grain size distribution:
\bea
I_{\star,w}^{\|,\perp}(\alpha,\psi)= \int da 4\pi a^{2}(dn/da)I_{\star,w}^{\|,\perp}(\alpha,\psi).\label{eq:Istar}
\ena

\subsection{Numerical Method and Results} \label{sec:busmodule}
We make use of ergodicity of the grain dynamical system to numerically compute the emission intensity given by Equation (\ref{eq:Istar_uv}). Basically, we can replace the ensemble average (i.e., over the angular distribution $f_{J}(\bJ)dJ_{x}dJ_{y}dJ_{z}$) by time average over all possible orientations and values of the grain angular momentum (ergodic theory). Thus, the emission intensity can be calculated as
\bea
I_{\star,w}^{\|,\perp}=\int f_{J}(\bJ)d\bJ\times\mathcal{I}_{w}^{\|,\perp} = \frac{1}{N}\sum_{\lbrace J,\beta,\varphi\rbrace_{i};i=1}^{i=N}\mathcal{I}_{w}^{\|,\perp}(J_{i},\beta_{i},\varphi_{i}),~~\label{eq:Iw_num}
\ena
where $\mathcal{I}$ denotes the entire term after $f_{J}(\bJ)d\bJ$ in Equation (\ref{eq:Istar_uv}). The polarization degree is then calculated by Equation (\ref{eq:pol_a}). For the case in which the magnetic field lies in the plane of the sky, we have $\zeta=\pi/2$. When the magnetic field is directed along the radiation direction, $\xi\equiv \beta$ and $\phi\equiv \varphi$. In this case, the fast Larmor precession allows averaging over the azimuthal angle $\varphi$ of $\bJ$ around the illumination direction $\zhat$.

\begin{figure*}
\centering
\includegraphics[width=0.49\textwidth]{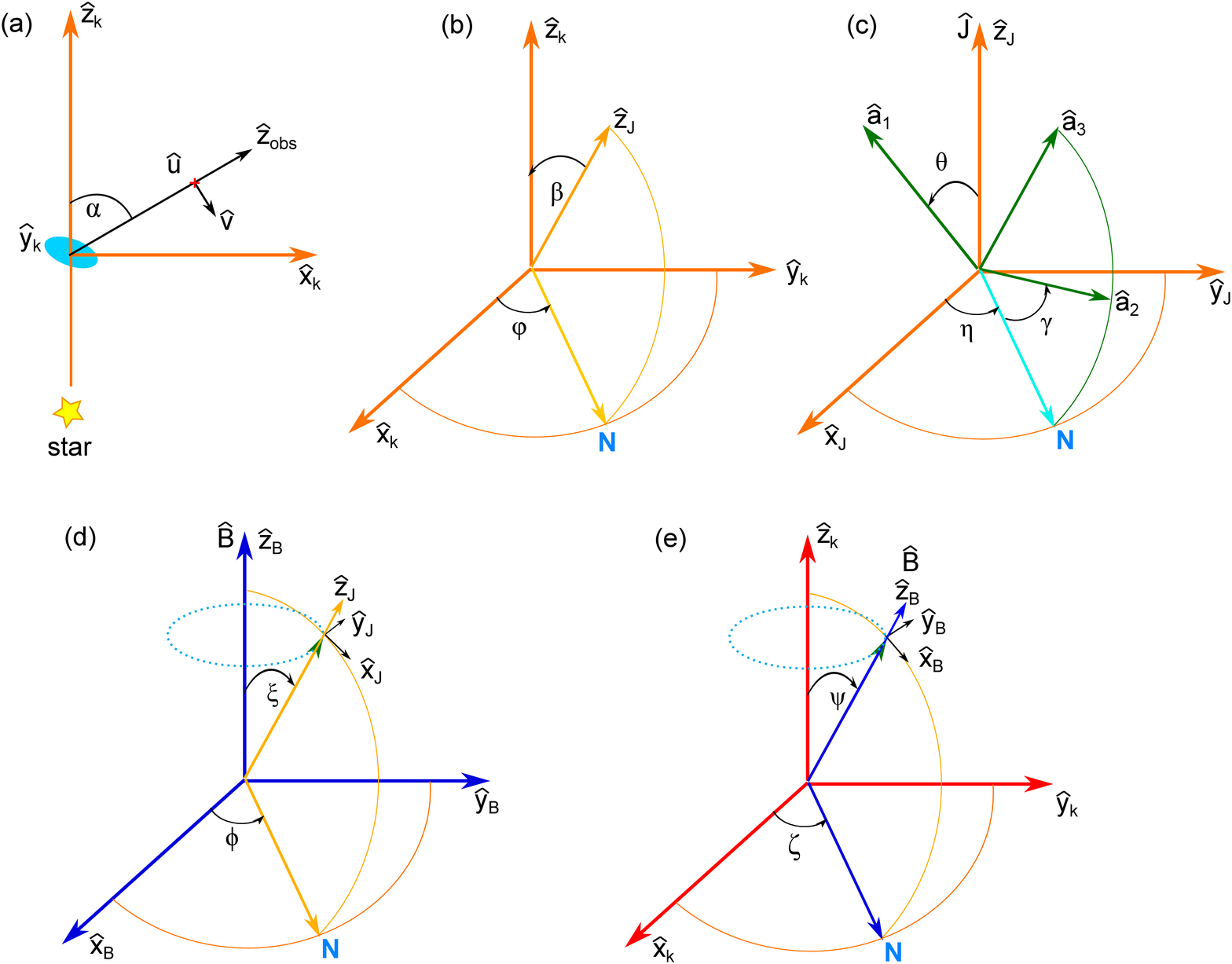}
\includegraphics[width=0.49\textwidth]{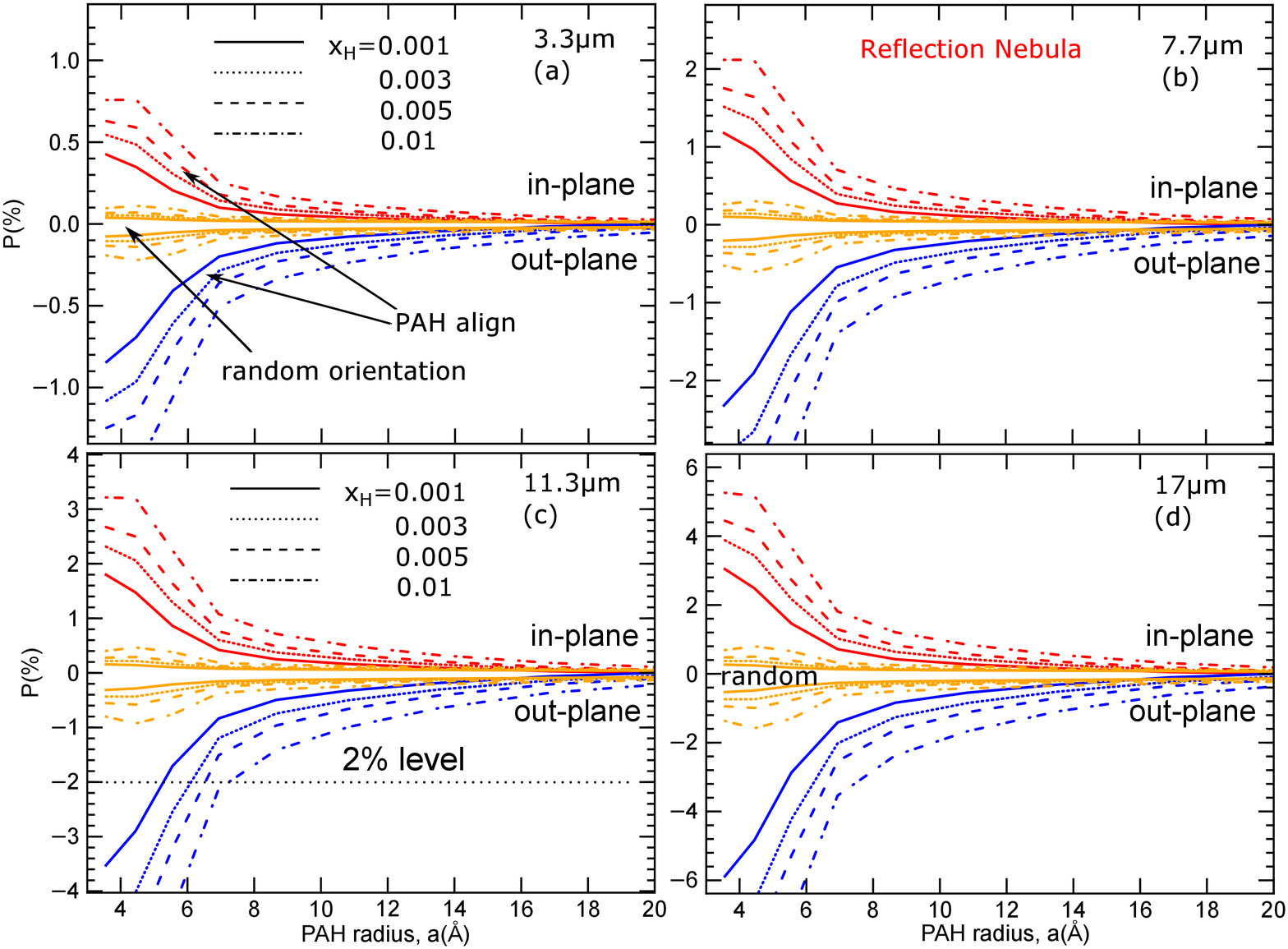}
\caption{{\bf Left panel}: Coordinate systems used in our calculations: (a) the star-molecule-observer system where $\zhat_{k}$ and $\zhat_{\rm obs}$ denote the illumination direction and the emission direction toward the observer; (b) orientation of the PAH angular momentum relative to the incident direction $\zhat_{k}$; (c) orientation of the PAH molecule in the reference frame defined by the angular momentum $\bJ$; (d) orientation of $\bJ$ in the magnetic field system; (e) orientation of the magnetic field in the reference system defined by the stellar incident radiation.
{\bf Right panel}: Polarization degree vs. PAH radius for selected PAH emission features for the RN conditions. Four different levels of gas ionization fraction $x_{\H}$ are considered. The polarization degree increases toward smaller $a$ as a result of the increase of the degree of external alignment.}
\label{fig:figures}
\end{figure*}

Our calculations show that the polarization is strongest for the reflection nebula (RN) conditions, whereas the polarization is predicted to be rather low for the diffuse ISM. Figure \ref{fig:figures}(right panel) shows the polarization degree predicted for the different emission features as a function of $a$. The polarization level tends to rapidly increase with decreasing $a$ for $a<10$\AA, whereas the polarization obtained for the case of random angular momentum increases rather slowly. The polarization level increases with the gas ionization fraction $x_{\H}$, as expected from the enhanced alignment due to stronger rotational excitation by ion collisions.

\section{Discussion and Summary}\label{sec:discus}

To date, observational studies of polarized PAH emission are still limited. Early observations by \cite{1988A&A...196..252S} report a tentative detection of polarized PAH emission from the Orion ionization front, with $p\sim 0.86\pm 0.28\%$ for 3.3 $\mu$m. Recently, a clearly detection of the polarization at 11.3$\mu$m from the MWC 1080 nebula is reported in \cite{Zhang:2017ea}. The measured polarization of $\sim 1.9 \pm 0.2 \%$ can be successfully explained when PAHs are aligned with the magnetic field to a degree of ~ $10\%$. Such alignment degree is naturally produced by resonance paramagnetic relaxation. Moreover, the analysis in \cite{Zhang:2017ea} found that the polarization direction is aligned with the ambient magnetic field. Thus, both polarization degree and direction are consistent with the prediction by alignment of grains with the magnetic field (\citealt{2017ApJ...838..112H}). Moreover, starlight polarization reveals a UV polarization bump from 2 stars (HD 197770 and HD 147933-4), and inverse modeling by \cite{2013ApJ...779..152H} indicates that PAHs are weakly aligned in the ISM toward one of these stars.

\cite{Hoang:2017vl} studied the cross-section alignment of PAHs by anisotropic radiation field and found that its efficiency is considerable, but the alignment by resonance relaxation is still dominant for strong magnetic fields.\footnote{\cite{2016ApJ...831...59D} suggested that the alignment of small PAHs is suppressed by the quantum effect. Yet the size below which the alignment is suppressed is not known for the reflection conditions.}

The alignment of PAHs as demonstrated through the detected polarization has two important implications. First, the alignment of PAHs with the magnetic field potentially opens a new window into studying magnetic fields via mid-IR polarization of PAH emission. For instance, polarized PAH emission would be most useful for tracing magnetic fields in the environments where strong PAH emission features are observed, such as RN, and circumstellar disks around Herbig Ae/Be stars \citep{2004A&A...427..179H} and T-Tauri stars.  Second, due to alignment, spinning PAH emission is polarized with the degree of a few percents for the diffuse ISM and up to $20\%$ at frequencies above 100 GHz for the RN \citep{Hoang:2017vl}. The polarization of spinning dust emission poses a challenge for the CMB B-mode detection.

\subsection*{Acknowledgments}
We thank the support by the Basic Science Research Program through the National Research Foundation of Korea (NRF), funded by the Ministry of Education (2017R1D1A1B03035359).




\end{document}